\documentclass{PoS}

\title{SS433, microquasars, and other transients}

\ShortTitle{SS433 and other transients}

\author{\speaker{Zsolt Paragi}\\
        Joint Institute for VLBI in Europe (JIVE), Postbus 2, 7990 AA Dwingeloo, The Netherlands\\
        E-mail: \email{zparagi@jive.nl}}

\author{Ren\'e Vermeulen\\
        Netherlands Institute for Radio Astronomy (ASTRON), Postbus 2, 7990 AA Dwingeloo, The Netherlands\\
        E-mail: \email{rvermeulen@astron.nl}}

\author{Ralph Spencer\thanks{This paper is dedicated to Prof. Istv\'an Fejes (1932-2011), 
enthusiastic researcher of SS433, who established radio astronomy (and especially VLBI) in Hungary.
e-VLBI research infrastructure in Europe is supported by the European Union's 
Seventh Framework Programme (FP7/2007-2013) under grant agreement RI-261525 NEXPReS. 
The EVN is a joint facility of European, Chinese, South African and other radio astronomy institutes 
funded by their national research councils.}\\
        Jodrell Bank Centre for Astrophysics The University of Manchester Oxford Rd Manchester M13 9PL UK\\
        E-mail: \email{ralph.spencer@manchester.ac.uk}}

\abstract{X-ray binaries have been an important key in understanding the jet-disc symbiosis 
          in accreting black holes on all mass scales, from stellar-mass to supermassive black holes. 
          SS433 was the first Galactic XRB that has been extensively studied in the radio regime.
          The radio properties, including the highest angular resolution data can now be better 
          understood in the framework for accretion disc state transitions that is observed in
          microquasars (black hole X-ray binary systems). SS433 remains unique in various ways to 
          date, and there is still much to learn about black hole accretion phenomena. In the meantime,
          the electronic very long baseline (e-VLBI) developments at the European VLBI Network (EVN) 
          has allowed us to study microquasars and other transients at milliarcsecond resolutions
          more flexibly than was possible before. Even more new opportunities will arise as the 
          SKA pathfinders become operational. 
          }

\FullConference{Resolving the Sky - Radio Interferometry: Past, Present and Future -RTS2012\\
		April 17-20, 2012\\
		Manchester, UK}

\begin{document}

\section{The first radio-jet X-ray binary}

A very unusual star received great attention at the end of 1970s which showed peculiar
H$\alpha$ emission lines: SS433, an eclipsing binary system that had counterparts 
in the X-ray and radio domain as well. The spectral lines showed high and variable
Doppler-shifts between -30000 and 50000 km/s, and these were soon interpreted as  
signatures of jets precessing with a period of about 164 days [for an early review,
see \pos{1}]. High resolution radio observation in 1979 with MERLIN indeed showed an 
elongated structure on arcsecond scale [\pos{2}], and the same year, compact structure
on milliarcsecond scales was reported following the first VLBI detection with a 3-station
array of Effelsberg, Onsala, and the Westerbork Synthesis Radio Telescope [\pos{3}].
Just in 1979, a dozen of Nature papers were published on the source, and the excitement
about this fascinating object certainly played a great role in the forming of the
European VLBI Network (EVN) in 1980. 

And there was a lot to discover. The highest angular resolution images of the precessing
radio beams allowed an accurate measurement of the source distance to 5$\pm$0.3~kpc
[e.g. \pos{4}]\footnote{Note that measuring the distances of transient Galactic radio sources
remains a big challenge to date; the uncertain distances and jet viewing angles still limit 
our ability to determine the jet Lorentz-factors accurately in black hole XRB.}
Due to the variable jet viewing angle and mildly relativistic speeds of 0.26c, it became 
possible to test the Doppler-beaming paradigm for relativistic jets as well [\pos{5}]. 
The source showed flaring episodes every $\sim$400~days. There were two types of flares
identified, one showing a delay with observing frequency, typical of synchrotron self-absorbed
ejecta, the other peaked simultaneously at all frequencies --the origin of the flares
were not known [\pos{6}]. Besides, there was a drop in flux density observed before the flares.
There were two VLBI monitoring campaigns organized in 1985 and 
1987, both lasting 6 days, with 2 day separation in between the observations. These were
really monster VLBI campaigns at the time, using practically the whole supplies of 
EVN recording tapes, weighing over a tonne. But the results were spectacular. There were
a series of moving ejecta detected that originated in an elongated core, and later 
brightened up at an angular distance of 40--60 mas (a brightening zone) -- these two regions
were responsible for the two different types of flares. The 1987 campaign coincided with 
a major flare by chance, and a series of spectacular images were produced. The Doppler-beaming
effect was confirmed both in the permanent core, as well as the in the moving ejecta, in 
agreement with the measured proper motion of 0.26c [\pos{7}].

   \begin{figure*}
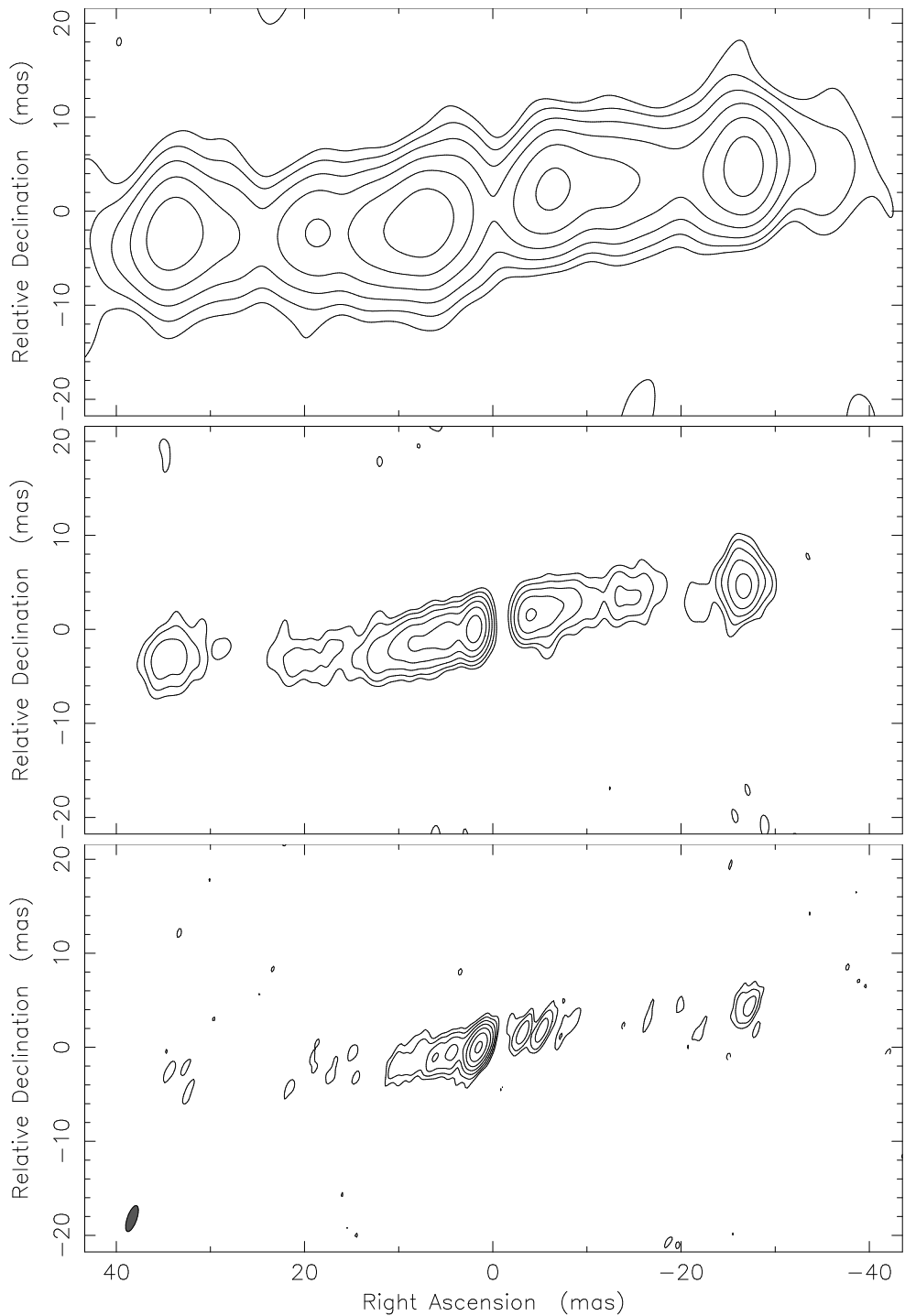

   \centering
   \vspace{20pt}
   \includegraphics[bb=33 295 582 550,clip,angle=0,width=13cm]{innerjet_18cm.ps}
   \includegraphics[bb=33 295 582 550,clip,angle=0,width=13cm]{innerjet_6cm.ps}
   \includegraphics[bb=33 260 582 550,clip,angle=0,width=13cm]{innerjet_2cm.ps}

   \caption{\label{compactjet} VLBA 1.6~GHz (top), 5~GHz (bottom) and 15~GHz (bottom) maps of 
   of SS433 on the same scale, demonstrating the multifrequency properties of the self-absorbed
   compact jets. 
   The compact object is
   located near the centre of these maps (found by fitting the kinematic model to the ejecta).
   The gap between the observed "radio cores" closes at higher frequencies, this effect is
   known in AGN as frequency dependent core-shift. The brightness assymetry at 1.6~GHz is 
   consistent with Doppler-beaming; the increasing assymetry at higher frequencies is likely 
   due to additional free-free absorption close to the central engine. 
   } 
    \end{figure*}

\section{The compact jets}

\begin{figure*}
   \centering
   \vspace{20pt}
   \includegraphics[bb=55 28 468 767,clip,angle=-90,width=15cm]{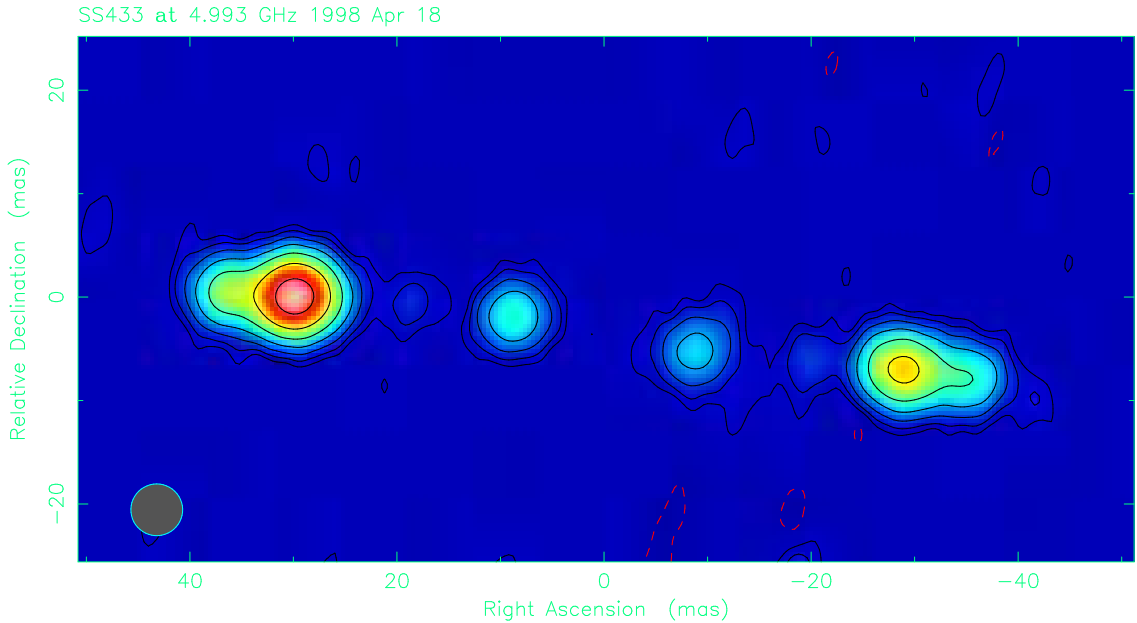}

   \caption{\label{nocore} 5~GHz VLBA map of SS433 during a strong flare on 18 April 1998.
   The core-region faded below detection, indicating that the continuous inner jets were 
   suppressed at the beginning of the flaring event. Quenching of radio emission from 
   the jets is now frequenctly observed in microquasars during accretion state transitions,
   when they enter into a soft-state, where the emission is dominated by the accretion disc.
   In SS433 there are no such accretion state changes observed, simply because the accretion 
   disc emission is hidden from us in the X-rays.
   } 
    \end{figure*}

Multifrequency imaging with the VLBA provided new insights to the nature of the
core region [\pos{8}]. The brightest peaks in the approaching and receding jets
have increasingly larger separation at lower frequencies. This is due to a change
in synchrotron opacity: the distance of the $\tau\sim1$ optical depth surface to 
the central engine is inversely proportional to the observing frequency in a 
self-absorbed compact jet. In active galactic nuclei (AGN) this is known as 
frequency dependent core-shift [\pos{9}]. Indeed, the compact object in SS433 lies
about midway between the eastern and western peaks of emission at 1.6~GHz (in the 
centre of images in Fig.\ref{compactjet}), not in the brightest "radio core" component. 
This was established by fitting the kinematic model to the moving ejecta in the system.
While this core-jet region has been observed most of the time in the system, there is
evidence for supression of the inner jet during (at least in some of the) flares [\pos{10}];
the temporal disappearance of the continuous jets before the flaring episode starts
can be therefore the cause of the flux density dips mentioned above (see Fig.\ref{nocore}). 

Compact jets are now frequently observed in microquasars in the low-hard X-ray state,
although resolved VLBI images are only available for a few of these. The activity cycle
of black hole XRB is understood as follows [\pos{11}]. The X-rays originate in a
comptonized corona at the jet base in the hard state. The accretion rate is a few percent 
Eddington, the jet Lorentz-factor is small ($<2$). As the accretion rate increases we
see an increasing level of hard X-rays and increasing radio emission. The X-rays later 
get gradually softer and the jet is suppressed (radio quenching). As the system enters
to the soft state (at which point most of the X-ray emission is thermal and dominated 
by the accretion disc), discrete radio ejecta form that are likely to be caused by shocks as
either the jet Lorentz factor or the mass flow rate increases. All of this had been 
observed in SS433 in the radio
before this unified picture for black hole XRB was established: the steady compact jets, 
radio quenching, and the formation of discrete ejecta. However, in case of SS433 the
X-ray emission from the accretion is not seen because it is almost completely hidden from 
us due to absorption and scattering by a dense circumstellar envelope.
Another difference is that in SS433 the accretion rate is highly super-Eddington all
of the time; still the similarities in the radio behaviour are remarkable. We have to 
note that the assumption of increasing Lorentz-factor during outbursts is certainly
not seen in SS433: the discrete, bright ejecta follow the usual kinematic model
with 0.26c.

\section{The equatorial outflow in SS433 and other black holes: driven by an accretion disk wind?}

Besides the steady inner jets and moving ejecta, there was another type of emission
seen in deep radio images, mainly at 1.6~GHz. This emitting region surrounds the
central source within $\sim$100 mas: some parts are occasionally detected as distinct
features in VLBI images [\pos{8}], while adding shorter spacings reveals its more
diffuse nature [\pos{12}]. Comparing multi-epoch detections there is evidence for 
proper motion in this region outwards from the centre, indicating an equatorial 
outflow [\pos{13}]. Similar outflows in microquasars and AGN were not known at the
time. However, there is now evidence in the X-ray observations for accretion disc driven ultra-fast 
outflows in both radio-loud [\pos{14}] and radio-quiet [\pos{15}] AGN. These may have 
significant contribution to the AGN feedback. In stellar-mass black holes similar
outflows were found in the soft states [\pos{16}]; these winds may carry sufficient
mass and energy to quench the jet radio emission as observed during state transitions. 
The massive equatorial outflow in SS433 may help us to understand another phenomenon
as well. In 2001 it was proposed that X-ray emission in the supercritical accretion
disc is channeled towards to the polar regions. When looked at face-on, SS433-like systems
may appear as ultra-luminous X-ray sources (ULX) in nearby galaxies [\pos{17}]. Recently,
new evidence was found for highly scattered supercritical disc funnel radiation [\pos{18}].

\section{Transients and the e-EVN}

During the last 8 years, the real-time e-VLBI developments made the EVN more capable of 
observing transient radio sources. This is because observations are no longer limited by 
recording media, the performance of the telescopes can be checked immediately in ad-hoc
observations, and the scheduling has become more flexible as well. Responses to flaring
activity is now possible in dedicated target of opportunity observations. For example,
e-EVN ToO observations of SS433 [\pos{19}] further confirm that discrete ejecta during
flares travel with the same speed as measured in the regular jet, therefore there is
no sign of an increasing Lorentz-factor in the jet during accretion disc state transitions. 
The e-EVN has provided rapid imaging results on several well-known microquasars and
other variable or transient sources in recent years. It has proven to be especially 
useful to combine the e-EVN and VLBA for denser monitoring observations, in which the
former can provide initial detections, improved coordinates for new transients, and
pre-select nearby secondary calibrator sources for more reliable imaging [e.g. \pos{20}].
Most recent results include the identification of a new gamma-ray binary stellar system
[\pos{21}], and the possible localization of the origin of gamma-ray flaring activity
in the Crab Nebula [\pos{22}]. 

The EVN and MERLIN played an important role in the very beginning of black hole XRB and
transient research with the observations of SS433 at the end of 1970s and in the 1980s,
and this is the case still today. With the electronic upgrades to e-EVN and e-MERLIN,
new opportunities arise: a combination of these two networks will provide a unique range
of uv-spacings to image transient sources from mas to arcsecond scales as they expand.
In the next few years a great number of new transients will likely be detected by SKA
pathfinder instruments and precursors like LOFAR or WSRT APERTIF. The great sensitivity
and high resolving power provided by the EVN will be essential to localize and image 
these at milliarcsecond resolution.


\begin{thebibliography}{99}

\bibitem[1]{} 
B. Margon, 
\emph{Observations of SS 433},
\emph{Ann. Rev. Astron. Astrophys.\/} {\bf 22} (1984) 507

\bibitem[2]{}
R.\,E. Spencer
\emph{A radio jet in SS433},
\emph{Nature\/} {\bf 282} (1979) 483 

\bibitem[3]{}
R.\,T. Schilizzi, C.\,A. Norman, W. van Breugel, E. Hummel,
\emph{VLBI detection of SS433},
\emph{Astron. Astrophys.\/} {\bf 79} (1979) 26 

\bibitem[4]{}
I. Fejes,
\emph{SS 433 extended radio structure observed with the European VLBI Network},
\emph{Astron. Astrophys.\/} {\bf 168} (1986) 69

\bibitem[5]{}
I. Fejes,
\emph{Does SS 433 exhibit relativistic beaming?},
\emph{Astron. Astrophys.\/} {\bf 166} (1986) 23

\bibitem[6]{}
R.\,C. Vermeulen,
\emph{Multi-wavelength studies of SS433},
\emph{PhD thesis,\/} {Leiden University, Leiden} (1989) 

\bibitem[7]{}
R.\,C. Vermeulen, R.\,T. Schilizzi, R.\,E. Spencer, J.\,D. Romney, I. Fejes,
\emph{A series of VLBI images of SS433 during the outbursts in May/June 1987},
\emph{Astron. Astrophys.\/} {\bf 270} (1993) 177

\bibitem[8]{}
Z. Paragi, R.\,C. Vermeulen, I. Fejes, R.\/T. Schilizzi, R.\,E. Spencer, A.\,M. Stirling, 
\emph{The inner radio jet region and the complex environment of SS433},
\emph{Astron. Astrophys.\/} {\bf 348} (1999) 910

\bibitem[9]{}
A.\,P. Lobanov
\emph{Ultracompact jets in active galactic nuclei},
\emph{Astron. Astrophys.\/} {\bf 330} (1998) 79

\bibitem[10]{}
Z. Paragi, R.\,C. Vermeulen, I. Fejes, R.\,T. Schilizzi, R.\,E. Spencer, A.\,M. Stirling,
\emph{VLBA multi-frequency monitoring of SS433},
\emph{New Astron. Rev.\/} {\bf 43} (1999) 553

\bibitem[11]{}
R.\,P. Fender, T. Belloni, E. Gallo,
\emph{Towards a unified model for black hole X-ray binary jets},
\emph{Mon. Not. R. Astron. Soc.\/} {\bf 355} (2004) 1105

\bibitem[12]{}
K.\,M. Blundell, A.\,J. Mioduszewski, T.\,W.\,B. Muxlow, P. Podsiadlowski, M.\,P. Rupen,
\emph{Images of an Equatorial Outflow in SS433},
\emph{Astrophys. J.\/} {\bf 562} (2001) 79

\bibitem[13]{}
Z. Paragi, I. Fejes, R.\,C. Vermeulen, R.\/T. Schilizzi, R.\,E. Spencer, A.\,M. Stirling, 
\emph{The Equatorial Outflow of SS433},
\emph{Proc. EVN Symposium\/} (2002) 263

\bibitem[14]{}
F. Tombesi, R.\,M. Sambruna, J.\,N. Reeves, V. Braito, L. Ballo,J. 
Gofford, M. Cappi, R.\,F. Mushotzky,
\emph{Discovery of Ultra-fast Outflows in a sample of Broad-line Radio Galaxis Observed with Suzaku}
\emph{Astrophys. J.\/} {\bf 719} {2010} 700

\bibitem[15]{}
F. Tombesi, M. Cappi, J.\,N. Reeves, G.\,G.\,C. Palumbo, T. Yaqoob, V. Braito, M. Dadina,
\emph{Evidence for ultra-fast outflows in radio-quiet AGNs I. Detection and statistical incidence of Fe K-shell absorption lines}
\emph{Astron. Astrophys.\/} {\bf 521} {2010} 57

\bibitem[16]{}
G. Ponti, R.\,P. Fender, M.\,C. Begelman, R.\,J.\,H., Dunn, J. Neilsen, M. Coriat, 
\emph{Ubiquitous equatorial accretion disc winds in black hole soft states}
\emph{Mon. Not. R. Astron. Soc.\/} {\bf 422} (2012) 11

\bibitem[17]{}
S. Fabrika, A. Mescheryakov,
\emph{Face-on SS433 stars as a possible new type of extragalactic X-ray sources}
\emph{Proc. IAU Symposium 205\/} (2001) 268

\bibitem[18]{}
A. Medvedev, S. Fabrika,
\emph{Evidence of supercritical disc funnel radiation in X-ray spectra of SS433}
\emph{Mon. Not. R. Astron. Soc.\/} {\bf 402} (2010) 479

\bibitem[19]{}
V. Tudose, Z. Paragi, S. Thruskin, P. Soleri, R.\,P. Fender, M.\,A. Garrett, R.\,E. Spencer,
A. Rushton, P. Burgess, M. Kunert-Bajraszewska, E. Pazderski, K. Borkowski, R. Hammargren, 
M. Lindqvist, G. Maccaferri,
\emph{e-VLBI observations of SS433 in outburst}
\emph{Astron. Telegram} {\bf 1836} (2008) 1

\bibitem[20]{}
J. Yang, C. Brocksopp, S. Corbel, Z. Paragi, T. Tzioumis, R.\,P. Fender,
\emph{A decelerating jet observed by the EVN and VLBA in the X-ray transient XTE J1752$-$223}
\emph{Mon. Not. R. Astron. Soc.} {\bf 409} (2010) 64

\bibitem[21]{}
J. Mold\'on, M. Rib\'o, J.\,M. Paredes,
\emph{Revealing the extended radio emission from the gamma-ray binary HESS J0632+057}
\emph{Astron. Astrophys.} {\bf 533} (2011) 7

\bibitem[22]{}
A.\,P. Lobanov, D. Horns, T.\,W.\,B. Muxlow, 
\emph{VLBI imaging of a flare in the Crab nebula: more than just a spot}
\emph{Astron. Astrophys.} {\bf 533} (2011) 10

\end{thebibliography}
\end{document}